\documentstyle[prl,aps,epsf]{revtex}
\begin{document}
\draft
\twocolumn[\hsize\textwidth\columnwidth\hsize\csname @twocolumnfalse\endcsname

\title{Oxygen-stripes in La$_{0.5}$Ca$_{0.5}$MnO$_{3}$ from {\it ab initio} calculations}

%\author{V. Ferrari  and P. B. Littlewood}

\author{V. Ferrari, M. D. Towler  and P. B. Littlewood}

\address{Cavendish Laboratory, University of Cambridge, Madingley
Road, Cambridge CB3 0HE, UK}

\date{\today}

\maketitle

\begin{abstract}

{\small

We investigate the electronic, magnetic and orbital properties of
La$_{0.5}$Ca$_{0.5}$MnO$_{3}$ perovskite by means of an {\it ab
initio} electronic structure calculation  within the Hartree-Fock
 approximation. Using the experimental crystal
structure reported by Radaelli {\it et al.} [Phys. Rev B {\bf 55},
3015 (1997)], we find a charge-ordering
stripe-like ground state. The periodicity of the stripes, and 
the insulating CE-type magnetic structure are in agreement with neutron x-ray 
and electron diffraction experiments.
However, the detailed structure is more 
complex than that envisaged by simple models of charge and orbital 
order on Mn d-levels alone, and is better described as a charge-density
wave of oxygen holes, coupled to the Mn spin/orbital order.

}

\end{abstract}

\pacs{PACS numbers: 75.47.Lx, 75.10.-b, 71.27.+a, 71.45.Lr}

\vskip2pc]
\narrowtext

Understanding the physics displayed by manganite oxide compounds
R$_{1-x}$D$_{x}$MnO$_3$ (R=rare-earth atom, D=divalent
substituent)  has stimulated much experimental and theoretical
work \cite{reviews}. Changing the composition $x$, they show a
variety of phenomena, such as ferromagnetic (FM), anti-ferromagnetic (AF),
charge and orbital ordering (CO and OO) revealing that charge, spin and and
lattice degrees of freedom are closely interrelated.

In the present work, we concentrate on the half-doped case of
La$_{1-x}$Ca$_{x}$MnO$_3$ that has been extensively studied
\cite{exptshalf,Radaelli,chencheong,theohalf,popovic}. 
For $x=0$, LaMnO$_3$,
there is a collective Jahn Teller (JT) distortion of the structure
that gives rise to a splitting of the $e_g$ levels. As a
consequence of that, the structure is orthorhombic, the Mn
$e_g$-levels are singly occupied and all Mn ions are
Mn$^{3+}$. For the other extreme doping ($x=1$), CaMnO$_3$ is
pseudo-cubic as the $e_g$-levels are empty and Mn$^{4+}$ does not couple
to the lattice distortion. In LaMnO$_3$, the splitting  of the
 $e_g$ orbitals favors the cooperative JT distortions and the
appearance of OO. This fact allowed Goodenough
\cite{Goodenough} and Kugel and Khomskii \cite{KugelKhomskii} to
explain the magnetic structure of this compound. Their explanation assumed unpolarized O$^{2-}$ ions and was
based on the idea that magnetic ordering is dictated by the
orientation of the orbitals involved.

The regime of intermediate doping is complex, magnetically and 
structurally, but even at the level of the {\it local} electronic
structure there are several experiments that present 
contradictory results. Some of them claim a mixed
valence picture of Mn$^{3+}$ and Mn$^{4+}$ while others found
features that do not reconcile with this image. For example, by
doing Oxygen K-edge electron-energy-loss spectra, it was concluded
that carriers in La$_{1-x}$Sr$_{x}$MnO$_3$ (0$\leq x\leq$ 0.7) have a significant oxygen-$p$ hole
character \cite{oxygenhole}. 
 Some x-ray absorption studies at the Mn K-edge in Ca-doped 
LaMnO$_{3}$ revealed a picture that does not match with 
a mixture of Mn$^{3+}$ and Mn$^{4+}$ for intermediate dopings
\cite{MnKedge}. On the contrary, Tyson \emph{et al.} \cite{Tyson}
performed a Mn $K_{\beta}$ x-ray emission experiment in the same
compound  and found that their data was compatible with a
Mn$^{3+}$/Mn$^{4+}$ mixing  for intermediate compositions. Similar 
contradictions have been found
using  neutron diffraction \cite{carvajal}, photo-emission and x-ray 
spectroscopy techniques \cite{moreexperiments}. These
contradictory features motivate the need for a clarifying picture
regarding the valence state of the Mn ions in manganite systems. One 
of the goals of the present work is to provide an {\it
ab initio} quantum mechanical calculation  to investigate the
local electronic structure of the compound in order to establish
the nature of the charge carriers.

In this letter we report the theoretical finding of oxygen-stripes
in La$_{0.5}$Ca$_{0.5}$MnO$_{3}$ by means of an  {\em ab initio}
spin-unrestricted Hartree-Fock (HF) \cite{UHF} study \cite{pattersonhalf}.
Our results were obtained by means of the code CRYSTAL98 
\cite{crystal98} that uses
a few localized basis functions per atom to solve
self-consistently the HF equations. In the present work, the basis
set for the different atoms are those optimized for previous
calculations \cite{dovesi,basisset} and the crystal structure is taken
from the experimental work reported by Radaelli {\it et al}
\cite{Radaelli}. So, the present study takes into account the
actual JT distortions and the size of the atoms
involved. It is expected that the periodic HF approximation will
correctly describe the physics involved in this system as it has
been successful in describing other magnetic insulators as well as
strongly correlated materials \cite{UHF}.

\begin{figure}[t]
\centerline{\epsfxsize=8.5truecm \epsfbox{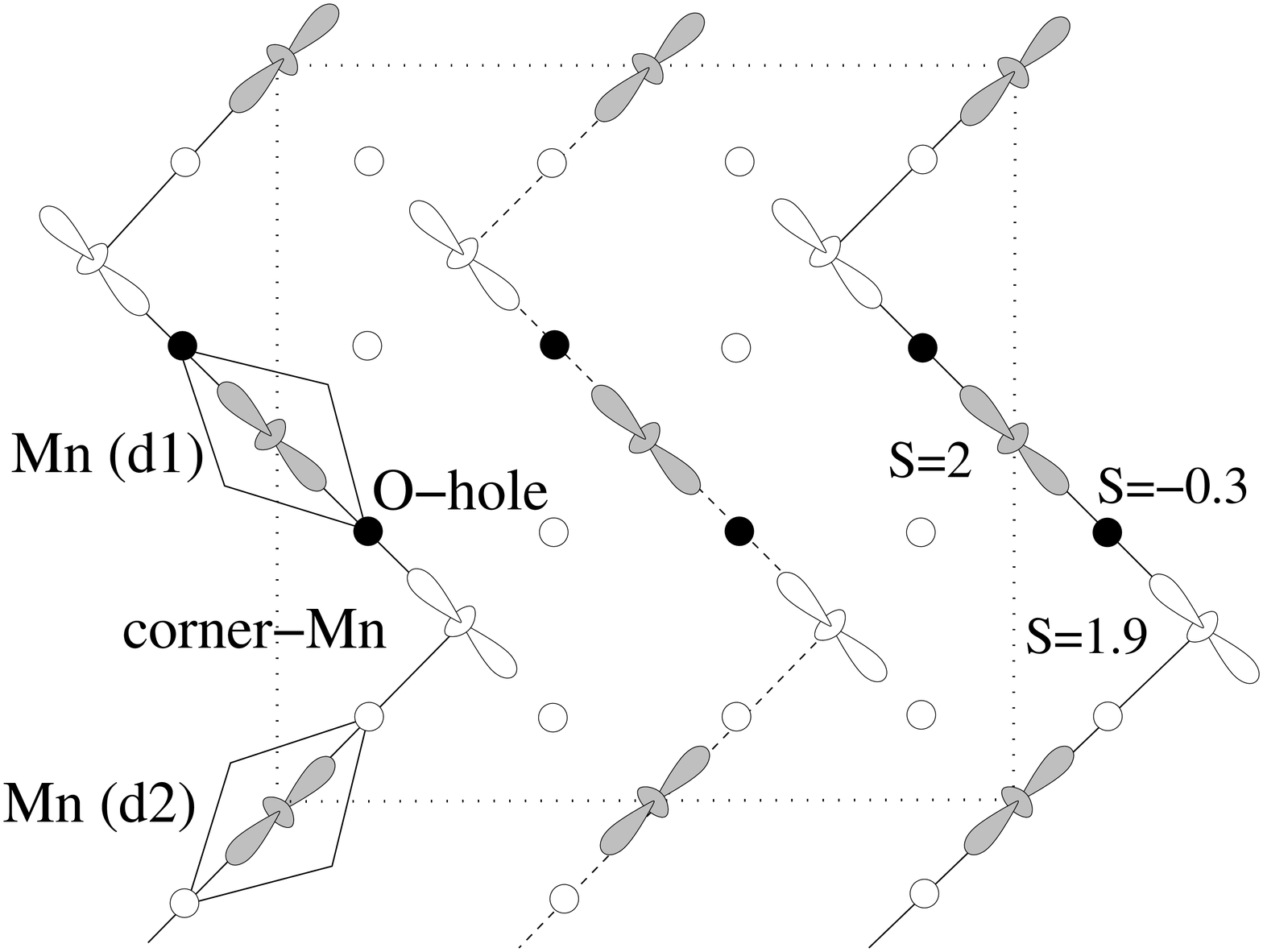} }
%\smallskip
\vspace{0.5cm}
\centerline{\epsfxsize=7.5truecm \epsfbox{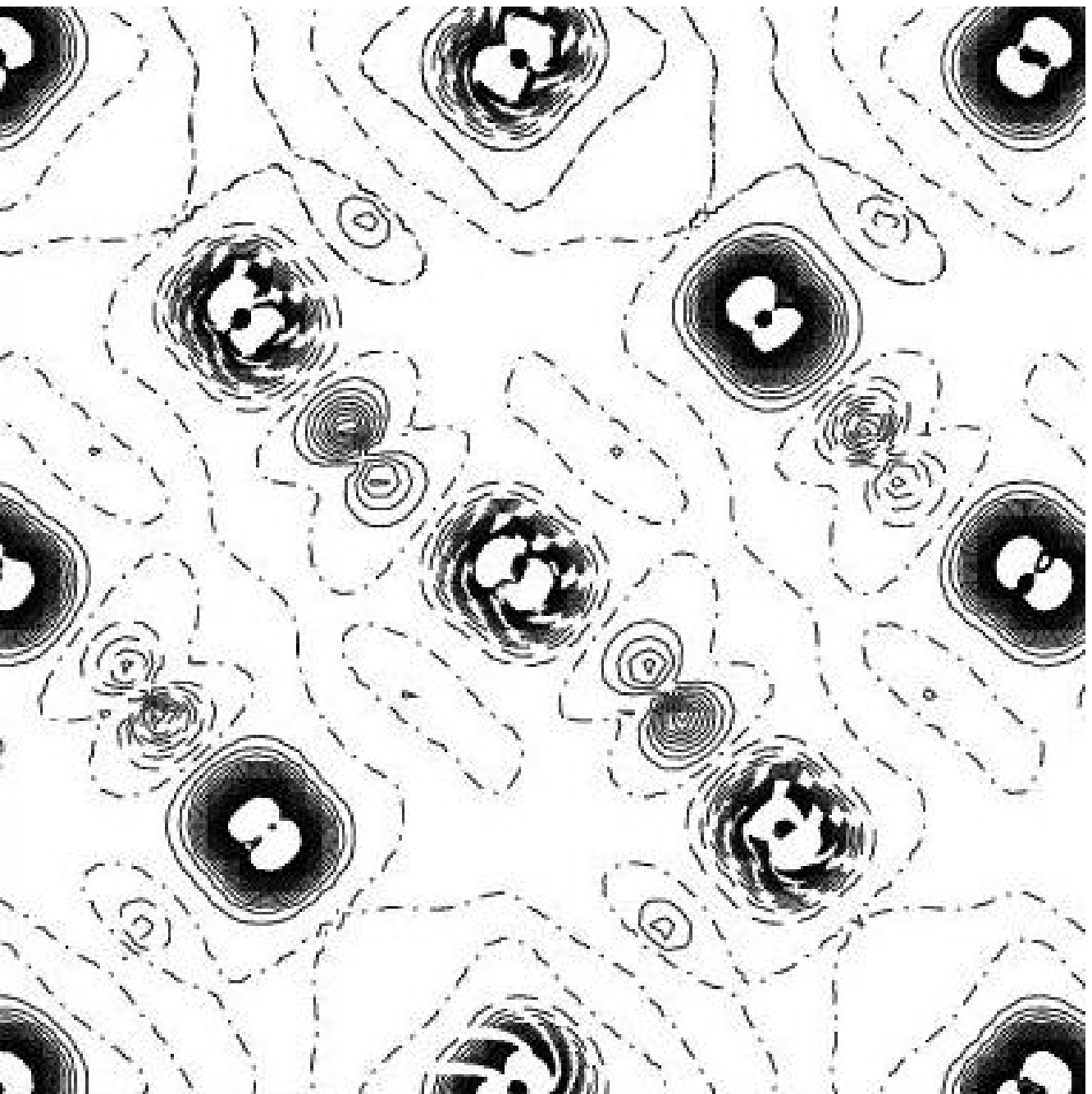}}
%\smallskip
\vspace{0.5cm}
\caption{{\it Top panel:} Sketch of the orbital ordering obtained
from {\it ab initio} HF calculations. Mn(d1)/Mn(d2) are the Mn in 
the distorted octahedra d1/d2. The ``corner-Mn'' is shown. Shading
indicates the spin polarization value (see Table \ref{table:mulliken}). Small circles represent 
the oxygens (black ones have spin-polarized holes).
Solid and dashed zigzag lines indicate oppositely spin polarized Mn species. 
The dotted line shows the unit cell used in all the calculations throughout this work.
{\it Bottom panel:} Spin-density map of a basal plane (corresponding to the unit cell in the sketch). Continuous,
dashed and dot-dashed isodensity curves correspond to positive, negative and zero values, respectively. The CE-type magnetic ordering and the spin-polarized oxygen holes are clearly seen. Adjacent basal planes are AFM coupled and display
stacking with identical Mn charge-orbital ordering.}
\label{sdmaps}
\end{figure}

The crystal structure for La$_{0.5}$Ca$_{0.5}$MnO$_{3}$ can be
indexed in space group $P2_{1}/m$ \cite{Radaelli} with 3
 inequivalent Mn atoms per unit cell. Two Mn ions
 are associated with JT distorted MnO$_6$ octahedra (that we will 
label as d1 and d2). Although in a similar environment of local O 
distortions, these two sites are not identical. The  third Mn  belongs to an almost
undistorted octahedra. The cell of the crystal structure has one axis
($b$) longer than the other two ones ($a,c$). We will call {\it basal
planes} the Mn-O planes perpendicular to the $b$ axis.
The magnetic ordering for this compound was coined CE-type \cite{wollan}. It consists of FM zigzag chains that are coupled AFM (see Fig.\ \ref{sdmaps})

As it was previously reported \cite{dovesi,kaplan} for
LaMnO$_3$ and CaMnO$_3$ within the HF approximation, Mulliken
population analysis (MPA) gives charges for Mn and O ions that
deviates substantially from the formal valence picture. In these
works, MPA results for LaMnO$_3$ give Mn$^{+2.2}$ and O$^{-1.7}$
to be compared with the formal values Mn$^{+3}$ and O$^{-2}$
whereas in CaMnO$_3$, Mn$^{+2.2}$ and O$^{-1.3}$ should be
contrasted with Mn$^{+4}$ and O$^{-2}$. However, the MPA for the
spin values do not show such deviation. In LaMnO$_3$, the Mn spin
was 1.98 \cite{kaplan} and in CaMnO$_3$, $\frac{3.2}{2}$
\cite{dovesi}; that is, very close to the formal valence spin
values of 2 and $\frac{3}{2}$, respectively. In both cases, the
spin for the O atoms was also close to the formal picture
being mainly spin unpolarized. The origin of the difference between 
charge and spin values is the strong hybridization between
Mn and O bands with the subsequent screening of Mn charge by electrons
in the surrounding O atoms. In the half doped case, we find that 
MPA also shows deviations from the formal valence picture with all
Mn having about the same charge. For the oxygens, almost all of them
have a formal charge close to  O$^{-1.6}$, but there are some holes
leading to O$^{-1.2}$, as we will detail further on.

 Regarding the magnetic properties of this system, we
found that the ground state is correctly predicted to
have the CE-type ordering in agreement with neutron experiments \cite{wollan,Radaelli}. The FM  state and the A-type AFM 
state both have energies that are above 
the CE-type state by at least 1.1 eV and 0.9 eV per unit cell, respectively \cite{note:energies}. Presumably, the FM 
and A-type states have similar energies because  the coupling between basal planes is weak.

We found that the ground state for the half doped case is almost doubly
degenerate. One of the solutions is
visualized in the spin-density map depicted in Fig.\ \ref{sdmaps}, that clearly shows that the Mn ions display OO. As an
aid to understand the map, we draw in the same figure, a sketch of
the corresponding orbital and crystal structure with shading
showing the value of the spin polarization (S). Some oxygens have spin-polarized holes (see Table \ref{table:mulliken}).  For the solution shown, holes are located on O atoms forming the long  bonds of one of the distorted
octahedra d1. The other (nearly degenerate) solution has oxygen holes instead along the long bonds of octahedra d2, with the corner-Mn orbital always in the direction of the O-hole \cite{note:solution}. So, within the 
HF approximation, not just the Mn
ions but also the oxygens play an important role. Indeed,
as  seen in Table \ref{table:mulliken}, charge modulation occurs 
principally on the oxygens rather than on Mn atoms.

\begin{table}[tb] \caption{Mulliken population values for the different atoms. The spin values
for the Mn d-orbitals are: $t_{2g}$=1.4, $d_{x^2-y^2}$=0.1, $d_{3z^2-r^2}$=0.5 for all of them \protect\footnotemark[1].} 
\label{table:mulliken}
\begin{tabular}[c]{c|c|c|c|c|c|c}
 & Mn(d1) & Mn(d2) & corner-Mn & O-hole \footnotemark[2] &La&Ca \\
%\hline
\hline
Charge & +2.1 & +2.2  &+2.1 & -1.2 &+2.9&+1.9 \\
Spin &  2.0 & 2.0 & 1.9 &-0.3 & 0.0&0.0\\
%\hline
\end{tabular} 
\small
\footnotemark[1] local system of reference on each Mn,  with z axis along the direction of the zigzag.
\footnotemark[2] The remaining oxygens have charge -1.6 and  are weakly polarized. 
\end{table}

\normalsize

The origin of the spin-polarized O-holes can be understood in terms
of the OO. For the analysis, we will consider only the atoms within a
 zigzag chain because the oxygens between chains are unpolarized
(which is consistent with the AFM coupling between adjacent Mn).
 As can be seen  in Fig.\ \ref{sdmaps}, the OO around Mn(d2) resembles
 LaMnO$_3$. Along the long bond of d2, we found no spin-polarized O-holes
which is consistent with Goodenough's ideas. However, around 
Mn(d1) the OO is different. In the long bond of d1, both Mn and O orbitals 
lie along the same line allowing for a transfer of charge from the O to 
the Mn, and rendering the O with a hole \cite{note:chargeflow}. 
Due to exchange, this O-hole has opposite spin to its two adjacent Mn ions.

\begin{figure}[t]
\centerline{\epsfxsize=3.truecm \epsfbox{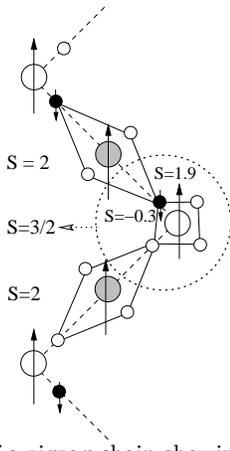} }
\caption{Detail of a zigzag chain showing the low-spin state in
manganites that is analogous to the Zhang-Rice singlet in
cuprates.}
\label{lowspin}
\end{figure}

As the spin-polarized oxygens are always AF
aligned with the neighboring Mn spin, the Mn-O complex forms a state analogous 
 to the Zhang-Rice singlet in the cuprates
\cite{zhangrice}: the corner-Mn  with the O-hole forms a low 
spin-state that gives a total S $\approx$ $\frac{3}{2}$ 
(see Fig.\ \ref{lowspin}). This
connects to the usual models of d-orbital occupancy alone,
where a hybrid Mn-O state stands in place of a Mn$^{4+}$ ion.
Since we find S $= 2$ for the remaining Mn atoms,
they match the conventional picture of Mn$^{3+}$ ions. So, although the
chemistry is quite complex involving holes on some oxygens, due to
the strong hybridization they are not independent degrees of
freedom and the system should be viewed in terms of hybridized
bands, possibly supporting the use of simplified single-band models.

The breaking of symmetry due to the localization of  holes on 
oxygens in only one of the distorted octahedra appears to highlight the 
very small symmetry breaking in the experimental crystal 
structure, where slightly different octahedra d1 and d2 have been 
reported \cite{Radaelli}. Namely, in the $a$-$c$ plane, the octahedra 
have long Mn-O distances of 2.07 and 2.06 \AA \, and short Mn-O distances of  
1.93 and 1.92 \AA, respectively. 
However, we performed a relaxation by increasing the  distance between the corner-Mn and O-hole
by 4 $\%$ and the  energy went down by 0.5eV per unit cell, suggesting that a full relaxation is 
needed. It is worth asking if a further refinement can be performed taking into 
account the features suggested by our results. In fact, a recent refinement for
 the structure of  Pr$_{0.6}$Ca$_{0.4}$MnO$_3$ \cite{carvajal} produces 
local ordering not  dissimilar to the results of our model.

We now compare our results with electron-microscopy
experiments (EM). As can be seen in Fig.\ \ref{sdmaps}, the CO and OO 
doubles the pseudo-cubic unit cell in one direction (vertical in Fig.\ \ref{sdmaps}) 
as reported in EM \cite{chencheong}. Note that the unit cell 
is further doubled in the horizontal direction due to the magnetic ordering
although this can not be explicitly seen in EM.
 It is important to note that either of the doubly degenerate solutions (either holes
on d1 or holes on d2) produce identical periodicity for the 
diffraction pattern. In a real sample, there will always be some perturbation to stabilize 
one of the solutions.

In experiments, although La and Ca are on average randomly distributed, little is known about local arrangement within a range of a unit cell.
We investigated this issue by checking  the influence of  different distributions for La and Ca within the planes just
above and below the basal ones.
We found that if the concentration of Ca is higher in the region above a particular octahedron (thus enhancing the symmetry breaking between d1 and d2), then holes locate in that octahedron  and the energy is lowered by 3.8 eV per unit cell. This result suggest that the local environment is important and should be taking into account in future refinements.

\vspace{-0.4cm}
\begin{figure}
\begin{center}
\centerline{\epsfxsize=10.truecm \epsfbox{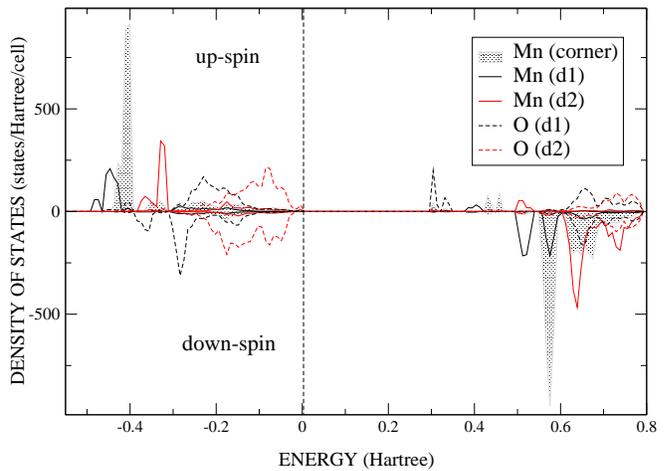} }
\end{center}
\vspace{-1cm}
\caption{DOS projected onto the atoms of
a zigzag chain. The dotted line marks the Fermi
energy. Indicated between parentheses is the octahedron to which each atom belongs.}
\label{DOS}
\end{figure}

Our picture is supported by an analysis of  the density of states (DOS) (Fig.\ \ref{DOS}).
In the valence band, the oxygens on d1 show a net down polarization (yet the other oxygens are almost
unpolarized) and the spin-polarized contribution of each type of Mn is 
essentially up. The top of the valence band is mainly populated by 
oxygen bands
showing the charge-transfer insulator character of this system.
The spin-polarized holes just at the bottom of the conduction band
 indicate that   the origin of the gap is probably  the localization 
of the charge produced by the oxygen-stripes that develop in the system, as a
 charge density wave (CDW).

In conclusion, we performed an {\it ab initio} study for the half
doped case of La$_{1-x}$Ca$_{x}$MnO$_{3}$ considering the
experimental structure \cite{Radaelli,wollan} and working within the HF
approximation. In contrast to the conventional model of CO that 
produces insulating behavior by the alternation 
of Mn$^{3+}$ and Mn$^{4+}$, our model yields an insulator because of 
ordering of O-holes. Nevertheless, the spin
character is exactly as predicted by the conventional models, because
each O-hole is bound into a local low spin state with the neighboring 
corner-Mn ion. The essential physics
 might be described with an effective Hamiltonian involving only Mn bands,
but this issue requires further investigation \cite{note:allaZhangRice}.
Interestingly,
we find that this hybridized  low-spin state further orders 
in a broken-symmetry CDW with AFM order.  
The unit cell and the periodicity of the oxygen-stripes are in agreement 
with neutron diffraction experiments. 
However, the O-hole density is substantial and not obviously compatible
with the very small difference between d1 and d2 octahedra reported in 
Ref.\cite{Radaelli}. Since the (La,Ca) disorder will favor locally 
broken symmetry states, it is likely that these short scale fluctuations will
restore the average symmetry of d1 and d2.
Future work should allow the structure to relax  within a
theoretical calculation  or to get a further refinement 
of the experimental  crystal structure for La$_{0.5}$Ca$_{0.5}$MnO$_{3}$.

The authors are indebted to Iberio de P. R. Moreira for 
helpful assistance with the CRYSTAL code and fruitful discussions. 
We are grateful for  valuable suggestions from M. Calder\'on, 
N. Mathur, D. Khomskii, V. Heine, and especially L. Colombi Ciacchi. V. F. acknowledges support from
the National Research Council of Argentina (CONICET) and
Fundaci\'on Antorchas (Argentina).  M. D. T. thanks Royal Society for support. P.B.L. acknowledges support
from EPSRC.

%%\begin{thebibliograph}{99}


\begin{references}
\bibitem{note:energies} These numbers can be affected by the different arrangement of La and Ca that we describe further on.
\bibitem{note:solution} All the results shown refer to the most stable solution (holes on d1) as the other one is analogous. Details of the calculation will be described elsewhere.
\bibitem{note:chargeflow} This charge flow from O to Mn atoms was also observed in density-functional calculations \cite{popovic}.
\bibitem{note:allaZhangRice} A theoretical treatment in the lines of 
Zhang-Rice theory \cite{zhangrice} with both Mn and O bands, should be considered.
\end{references}
\end{document}